\documentclass{hotnets22}
\usepackage{times}  
\usepackage[pdfpagelabels=false]{hyperref} 
\usepackage{graphicx}
\usepackage{textcomp}
\usepackage{booktabs} 
\usepackage{array}
\usepackage{titlesec}
\usepackage[table]{xcolor} 
\usepackage{fontawesome}
\usepackage[scaled=0.8]{beramono} 

\usepackage{wasysym}
\usepackage{comment}

\hypersetup{pdfstartview=FitH,pdfpagelayout=SinglePage}

\definecolor{darkblue}{rgb}{0,0,0.5}
\hypersetup{
	pdftitle={Tango or Square Dance? How Tightly Should we Integrate
Network Functionality in Browsers?},
	pdfauthor={A. Davidson, M. Frei, M. Gartner, H. Haddadi, J. S. Nieto, A. Perrig, P. Winter, and F. Wirz},
	pdfkeywords={},
	colorlinks=true,
	urlcolor=darkblue,
	linkcolor=darkblue,
	citecolor=darkblue,
}

\colorlet{tableheadcolor}{white} 

\colorlet{color1}{red!60!white}
\colorlet{color2}{blue!60!white}
\colorlet{color3}{green!80!black}
\colorlet{color4}{orange!60!white}
\definecolor{colorbackground}{rgb}{0.918,0.918,0.949} 
\colorlet{rowcolor1}{blue!70!green!10}
\colorlet{rowcolor2}{blue!77!green!15}
\colorlet{heatmapMin}{red!70}
\colorlet{heatmapMid}{orange!70}
\colorlet{heatmapMax}{orange!30}

\newcommand{\topline}{\arrayrulecolor{black}\specialrule{0.1em}{\abovetopsep}{0.5pt}%
    \arrayrulecolor{tableheadcolor}\specialrule{\belowrulesep}{0pt}{-3pt}%
    \arrayrulecolor{black}
}
\newcommand{\midline}{\arrayrulecolor{tableheadcolor}\specialrule{\aboverulesep}{-1pt}{0pt}%
    \arrayrulecolor{black}\specialrule{\lightrulewidth}{0pt}{0pt}%
    \arrayrulecolor{white}\specialrule{\belowrulesep}{0pt}{-3pt}%
    \arrayrulecolor{black}
}

\newcommand{\bottomline}{\arrayrulecolor{white}\specialrule{\aboverulesep}{0pt}{-2pt}%
    \arrayrulecolor{black}\specialrule{\heavyrulewidth}{0pt}{\belowbottomsep}}%

\ifdefined\final
\newcommand{\gennote}[3][blue]{}
\else
\newcommand{\gennote}[3][blue]{\textcolor{#1}{\ $\rule{8pt}{8pt}_{\textsf{\scshape\bfseries #2}}$ #3}}
\fi

\titlespacing{\section}{0pt}{*1.5}{*1.5}
\titlespacing{\subsection}{0pt}{*1.5}{*1.5}
\titlespacing{\subparagraph}{\parindent}{*1.5}{*1.5}

\makeatletter
\def\@listi{\leftmargin\leftmargini
    \parsep 1\p@ \@plus0\p@ \@minus\p@
    \topsep 2\p@   \@plus0\p@ \@minus\p@
    \itemsep1\p@ \@plus0\p@ \@minus\p@}
\let\@listI\@listi\@listi
\makeatother

\begin{document}

\title{Tango or Square Dance? How Tightly Should we Integrate Network Functionality in Browsers?\vspace{-1.5em}}

\author{
  {\rm
    A. Davidson\textsuperscript{$\bigstar$}\quad
  }
  {\rm
    M. Frei\textsuperscript{$\ddagger$}\quad
  }
  {\rm
    M. Gartner\textsuperscript{$\maltese$}\textsuperscript{*}\quad
  }
  {\rm
    H. Haddadi\textsuperscript{$\bigstar$}\textsuperscript{$\dagger$}\quad
  }
  {\rm
    A. Perrig\textsuperscript{$\ddagger$}\quad
  }
  {\rm
    J. Subirà Nieto\textsuperscript{$\ddagger$}\quad
  }
  {\rm
    P. Winter\textsuperscript{$\bigstar$}\quad
  }
  {\rm
    F. Wirz\textsuperscript{$\ddagger$}\thanks{Corresponding authors: \email{{\rm marten.gartner@ovgu.de}}, \email{{\rm wirzf@ethz.ch}}}
  }
  \\
  \textsuperscript{$\bigstar$}{\rm Brave Software}\qquad \textsuperscript{$\ddagger$}{\rm  ETH Zurich}\qquad 
  \textsuperscript{$\dagger$}{\rm  Imperial College London}\qquad
  \textsuperscript{$\maltese$}{\rm  OVGU Magdeburg}
}

\maketitle

\begin{abstract}
The question at which layer network functionality is presented or abstracted
remains a research challenge. Traditionally, network functionality was
either placed into the core network, middleboxes, or into the
operating system -- but recent developments have
expanded the design space to directly introduce functionality into the
application (and in particular into the browser) as a way to
expose it to the user.

Given the context of emerging path-aware networking technology, an
interesting question arises: which layer should handle the new
features? We argue that the browser is becoming a
powerful platform for network innovation, where even user-driven
properties can be implemented in an OS-agnostic fashion. We
demonstrate the feasibility of geo-fen\-ced browsing using a prototype browser extension,
realized by the SCION path-aware networking architecture, without introducing any significant performance overheads.
\end{abstract}

\section{Introduction}

With the emergence of path-aware networks (PAN)~\cite{pan2018}, the question
arises on how applications can harness the new network properties.  In
particular, Segment Routing (SR)~\cite{RFC8402} is emerging
in intra-domain environments, and SCION is becoming available in
several inter-domain locations~\cite{Chuat2022,conext2021deployment}.

As these architectures are seeing real-world deployment, new
opportunities emerge. In particular, PANs offer
multiple path options from which the end system can select from --
simultaneously also providing native inter-domain multipath.
Furthermore, path-aware architectures can decorate network paths
with additional information, such as latency, expected bandwidth, MTU,
traversed ASes, carbon footprint, etc. The combination of multiple
path options and per-path information enables exciting new
possibilities. For instance, network \textit{performance} can be
improved by selecting latency- or bandwidth-optimized paths, by
selecting a path with low loss,
or by knowing the accurate path MTU.
\textit{Communication quality} can also be improved through
selection of paths with low jitter, or through QoS offerings some
PAN architectures provide~\cite{conext2021colibri}.
PANs also enable enhanced \textit{privacy} through a
property referred to as geofencing, defined by fine-grained control over
which ASes are encountered on the forwarding path, similar to Alibi
routing~\cite{alibi2015}. Path decorations may also include
environmental, societal, and governance (\textit{ESG}) information,
such as the carbon footprint of the traversed ASes, giving rise to ESG-based path
selection. Finally, \textit{economic} aspects can be used for path
selection, with the obvious low\-est-cost routing, but for instance also enabling
selection of paths based on allied ASes.

Given these new opportunities for path selection, interesting
research questions arise:
\begin{itemize}
\item What entity should collect the path information? How is
  the information disseminated? How is this information authenticated
  (or where applicable certified)?
\item For the property classes of
  performance, quality, privacy, anonymity,
  ESG, and economics, at what layer \emph{can} or \emph{should} the path
  decision be made? Entirely ``within''
  the network? By the operating system (OS)? By the application? Or
  even by the user?
\item What are the interfaces between the network layers to convey
  path information (potentially all the way to the user)?
\end{itemize}

In this paper, we consider the general problem of which property
would be best addressed at what layer. In particular, we consider
network properties that one can implement in browsers~--~potentially
even in a user-driven manner. In that way, we leverage the browser to drive network innovation.

Browsers have already served as vehicles for network innovation, as in
the deployment of QUIC, which was a decisive force
to making UDP work well throughout the Internet.
QUIC is also deliberately built in user-space to enable further evolution to be driven from the applications, be it on the browser or web server side.
Another example is the pervasive deployment of VPN technology, which is
directly built into many browsers to provide improved privacy and
anonymity properties. The rapid deployment of DNS-over-HTTPS (DoH) and
DNS-over-TLS (DoT) constitute another example.  Being deployed in browsers, DoH and DoT bypass the operating system's DNS resolver.

Browser-based integration of network functionality offers several
compelling benefits. Integrating new network
functionality in the network or the OS comes with near insurmountable
challenges caused by highly heterogeneous infrastructure and
long update cycles.
As a small number of browsers see high penetration~\cite{web-browser-market-share}, and as browsers run on a variety of platforms, new functionality can be
disseminated to a spectrum of users with relative ease.

Another important benefit relates to usability. Many
brow\-sers update automatically (requiring minimal user intervention),
making it possible to disseminate new features rapidly and
comprehensively~\cite{chromium-release-cycle}.

Another aspect of this usability is that a browser integration
can design interfaces for directly interacting with the user itself if
needed. The Brave browser provides a concrete motivating example for these
considerations, as it provides a tight integration with the Tor
network: a user can simply open a browser window for anonymous
communication, avoiding a manual installation of Tor~\cite{brave-tor}.

To make our discussion more concrete, we leverage the benefits of tight browser integration to instantiate the PAN
architecture with SCION~\cite{Chuat2022}, which is deployed as a
prod\-uction-ready, next-generation network architecture currently operated by 12 ISPs.
We consider the geofencing network property in more detail, and
present an implementation in the Brave browser. Our approach demonstrates
that browser vendors are a powerful ally when deploying new networking
functionality such as PAN.

\section{Which Layer Should Make Path Decisions?}
\label{sec:which-layer}

Given the exciting new properties that PAN architectures offer, we
discuss in this section which layer should best make path
decisions.

Table~\ref{tab:properties} lists the set of properties we consider.

\begin{table}[t]
\tiny
    \centering
    \caption{Different properties enabled by path-aware networking. The
       {\CIRCLE} marks indicate that the layer can meaningfully select a path
       based on the property. In contrast, the {\Circle} marks indicate that the layer would not be the appropriate
       place to perform the path selection. A {\LEFTcircle} mark shows that no
       particular benefits are expected.
      }
    \label{tab:properties}
    \setlength{\tabcolsep}{5pt}
    \rowcolors{2}{rowcolor1}{rowcolor2}
    \resizebox{\linewidth}{!}{
    \begin{tabular}{l c c c}
        \topline
        \textbf{Property} & \textbf{OS} & \textbf{App} & \textbf{User}\\
        \midline
       \multicolumn{4}{l}{\textbf{Performance properties}}\\
       Low latency & \CIRCLE & \LEFTcircle & \LEFTcircle \\
  	   Loss rate & \CIRCLE & \CIRCLE & \Circle \\
       Path MTU information & \CIRCLE & \LEFTcircle & \Circle \\
	   Bandwidth & \CIRCLE & \CIRCLE & \LEFTcircle \\
       \hline

       \multicolumn{4}{l}{\textbf{Quality properties}}\\
	   QoS & \CIRCLE & \CIRCLE & \LEFTcircle \\
	   Jitter optimization & \CIRCLE & \CIRCLE & \LEFTcircle \\
       \hline

       \multicolumn{4}{l}{\textbf{Privacy / Anonymity}}\\
	   Geofencing (Alibi routing) &  \LEFTcircle & \CIRCLE & \CIRCLE \\
	   Onion routing &  \LEFTcircle & \CIRCLE & \CIRCLE \\
       \hline

       \multicolumn{4}{l}{\textbf{ESG Routing}}\\
	   Carbon footprint reduction & \LEFTcircle & \CIRCLE & \CIRCLE \\
	   Ethical routing & \LEFTcircle & \LEFTcircle & \CIRCLE \\
       \hline

       \multicolumn{4}{l}{\textbf{Economic aspects}}\\
	   Allied AS routing & \LEFTcircle & \CIRCLE & \CIRCLE \\
	   Price optimization & \CIRCLE & \CIRCLE & \CIRCLE \\
       \bottomline
    \end{tabular}
    }
    \vspace{-3em}
\end{table}

The network layer implements PAN mechanisms in both the control and
data plane. For most properties,
the control plane aggregates the required information and decorates
the path segments that are established. In the SCION context, the
path-segment construction beacons sent from AS to AS, iteratively
accumulate information during construction~\cite{Chuat2022} -- similar
to a BGP update traversing the Internet. The created path segments are
then disseminated through a path server infrastructure, along with the
additional information. End hosts fetching path segments thus receive
the fully decorated paths containing all added information.

We seek to address the following question: at what layer should path selection
take place? As the end host selects the end-to-end path from a set offered by
the network, the network layer has limited discretion about which path the
packet traverses.
Instead, the network layer relies on enforcing policies regarding which paths
are created and disseminated, and how much bandwidth can be obtained in the data
plane.

Consequently, the ultimate decision point for the path selection is at
the end host, which can choose from a set of offered paths. Depending
on the network topology, SCION can offer on the order of dozens to
even over a hundred potential paths through the combination of
different path segments. Such a large number of path choices offer a
meaningful way for multi-criteria end-to-end path optimization.

The question thus remains at what layer path
selection should be implemented. We see three broad possibilities: OS, application, and
user. Table~\ref{tab:properties} lists various PAN properties
along with the perceived best locus of decision. The OS networking stack can
select the path based on performance or quality properties:
low-latency and high-bandwidth connections clearly provide a good user
experience, especially if that connectivity is available at a low price.
However, for properties such as privacy, anonymity, or ESG (environment,
society, governance) routing, the OS generally lacks context to determine that
traffic is privacy sensitive, or how much performance the user is willing to
trade for better ESG metrics.
Conversely, the user cannot make an informed decision for some metrics.
Metrics such as loss and MTU get abstracted by lower layers, since they are
directly impacted by their interactions with the transport layer and OS.

With a path-based network API, the application can perform application-specific
path optimizations, such as selecting low-latency paths for the voice
channel of conferencing applications, or low-loss paths for IoT
command-and-control channels, or anonymity for medical web sites.

An interesting observation of these considerations is that for some
properties the user context is decisive, as an application can hardly
figure out automatically for which destinations CO$_2$ optimization is
desired, and when geofencing (restricted to which areas) should be used.

\section{Network Innovation in the Browser}
\label{sec:network-innovations-in-browser}

Section~\ref{sec:which-layer} indicates that operating PAN architectures at the application layer provides advantages over operating at the OS layer, which raises the question: \emph{in which applications should such architectures run}? 

\subsection{Network-specific Browsing Tabs}

A natural answer to this question is to develop solutions for web browsers. Browsers are the primary me\-di\-um by which people interact with the Internet. In 2021, five billion individuals used a web browser as part of their desktop or mobile phone usage~--~with 3.2 billion of those using Google Chrome~\cite{web-browser-stats}. Developing PAN solutions in the browser enables desktop and mobile users to immediately benefit from the associated networking advantages.

One of the main benefits of deploying new technologies within a web browser is that this minimizes the amount of configuration and installation that is required from novice users. In particular, many users are already familiar with the browser that they use, and such technologies can typically be integrated without any additional setup. The absence of any additional configuration also ensures that comprehensive security settings can be applied by default without any user intervention. Overall, this allows providing the strongest possible privacy and security guarantees for all users.

Integrations of this nature have already been demonstrated, such as in the Brave browser's integration of Tor-powered browsing windows~\cite{brave-tor,USENIX:DinMatSyv04}. This allows users to take advantage of Tor's anonymity guarantees simply by opening a new window, without manual installation of the Tor daemon. Usage data from June 2022 suggests that around 30\% of Brave users have opened a Tor-enabled window at some point~\cite{brave-internal-stats}, highlighting the popularity of this feature. The Brave browser reached 60M monthly active users in 2022~\cite{brave-mau}, which translates to around 20M users that have used this functionality. With Tor metrics indicating that around 3M direct connections are currently made per day~\cite{tor-usage}, we cannot underestimate the power of such browser-based integrations in bringing advanced networking technologies to much wider groups of individuals.

\subsection{High-level Architecture and Challenges}

To implement PAN functionality within browsers, there are two widely accepted alternatives: 1)~for browsers whose source code is freely available (e.g., Chromium, Firefox, and Brave), one can add new functionality directly into the brows\-er; and 2)~providing new functionality as \emph{browser extensions} using the widely adopted WebExtensions frame\-work -- supported by almost all browsers. Both approaches come with their own challenges.

Firstly, writing functionality directly into browser source code requires expertise with complex browser internals; replicating patch sets across multiple code bases (for supporting multiple browsers); and the respective browser maintainers need to accept the patch set, which can be a substantial political challenge. Using browser extensions allows wrapping functionality into self-contained units of software that can be distributed independently of browser-related release cycles, without requiring expert knowledge of browser internals. However, a main drawback of extensions is that their control over the browser is limited by design, and users must also explicitly install them.

Considering these trade-offs, we have chosen to implement an initial proof-of-concept PAN-based application in the WebExtensions framework. This gives us the advantage of building a prototype that can be demonstrated to work in today's browsers. We show in Section~\ref{sec:pan-browsing} that we can achieve a coherent PAN-based networking application using the various WebExtensions APIs, and that these can be integrated seamlessly into Chromium-based browsers like Brave. We anticipate that our results 
will encourage browser makers to work towards integrating more networking-based technologies directly into browsers in the long-term.
\section{Path-Aware Networking in the Browser}
\label{sec:pan-browsing}

SCION is a path-aware inter-domain network architecture, organizing autonomous systems (ASes) into \emph{isolation domains} (ISDs) which define local trust roots for SCION's control plane public-key infrastructure (PKI)~\cite{Chuat2022}.
These ISDs typically comprise ASes sharing a common legal framework and are thus bounded geographically by a country or by a compatible political entity.

Paths are discovered with an announcement process called \emph{beaconing}. The path information announced by the individual ASes includes various metatadata items and is authenticated based on the control-plane PKI.
For each AS hop on a path, this metadata consists of the ISD and AS numbers, and various optional items such as MTU, latency, bandwidth, geographic coordinates, or data on power efficiency and carbon emissions of the AS's infrastructure.

\begin{figure}[t]
\includegraphics[width=\linewidth]{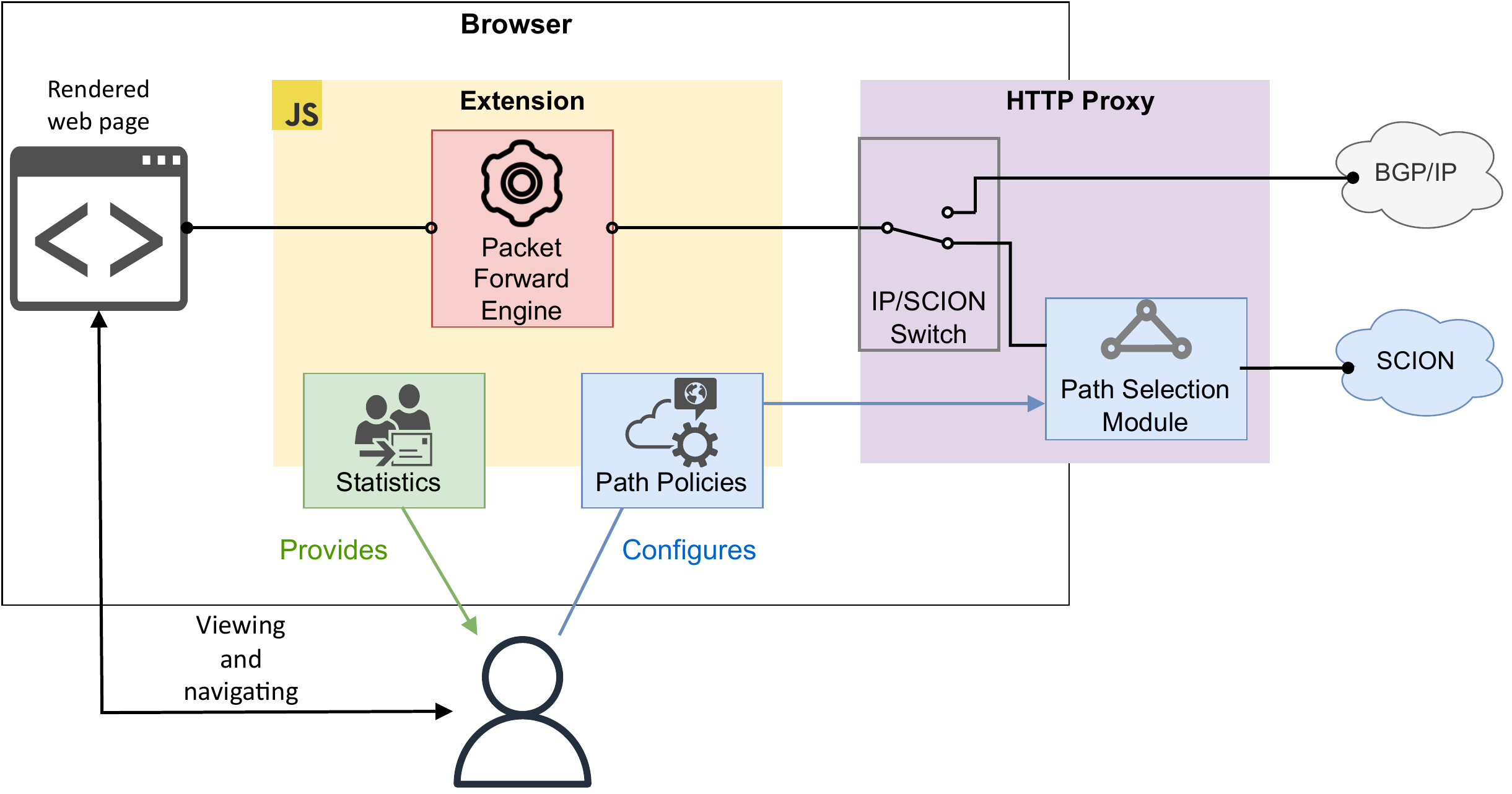}
\caption{PAN using SCION for browsers.}
\label{fig:architecture}
\end{figure}

SCION's path-awareness provides opportunities for applications and users to optimize data transport over the Internet. We propose a design for integrating PAN over SCION into existing browsers to bring path-aware networking to users, as depicted in Figure~\ref{fig:architecture}. The core of our design consists of two components with respective open-source implementations: A browser extension \cite{scion-browser-extensions} that intercepts requests initiated by the browser interacting with a HTTP proxy \cite{scion-apps-skip} that selects path(s) and adds a SCION packet header if needed. The path selection is performed depending on a set of rules configured by the user, called \emph{path policies}. Finally, statistics on path usage and performance of particular paths are provided as feedback to users.
In case the client or server lacks SCION connectivity, the browser falls back to loading the resources over IPv4/6.

\subsection{Path Policies and Geofencing}
To establish a connection to a remote host, a SCION application s the set of available candidate paths from the local AS path service, which include metadata added during beaconing.
The application then utilizes these path metadata to evaluate the path policy locally, filtering out any non-compliant paths.

Path policies are rules to filter the available SCION paths to a particular destination expressed by a dedicated \emph{Path Policy Language} (PPL) \cite{pathlanguage}. Based on this language, policies can be designed to
sort and select paths depending on specified criteria, such as bandwidth, latency or included hops.
Multiple policies can be combined for fine-grained configuration, e.g., optimizing the CO$_2$ footprint while excluding particular regions.
Note that the path policy remains on the user's device and does not need to be shared with any external services.
It may however still be possible to \emph{infer} a user's policy by observing their network traffic.

Geofencing is implemented by selecting paths that include or exclude ASes or ISDs.
Currently, we perform geofencing at the ISD-level.
We provide the user with an interface to block or allow entire ISDs.
Since ISDs are designed to cover independent regions or networks, we anticipate a balanced degree of customization, while keeping the number of options manageable. However, by integrating with the PPL, we provide the foundation for finer-grained geofencing.

\subsection{Partial Availability}

Our approach depends on the availability of SCION both on the client side as well as on the web server side. Even if SCION adoption for websites rises rapidly, many websites will likely remain unreachable via SCION in the near term.
In order to keep websites working despite this limited availability, and at the same time provide meaningful geofencing properties, we define two operational modes. In the default \emph{opportunistic mode}, SCION is used whenever available. Third-party resources that cannot be loaded over SCION are fetched over IPv4/6. An icon in the browser's UI indicates to the user whether all, some, or no parts of the website were fetched over SCION.
To provide stronger geofencing guarantees, the user can selectively enable \emph{strict mode}, e.g., for particularly sensitive websites, where all resources must be loaded over SCION. In this mode, websites may fail to load completely. For websites that are accessible over SCION, including any third-party resources, operators can indicate this including an HTTP response header \texttt{Strict-SCION}. Upon receiving this header, the browser enforces strict mode SCION for requests to the host from whom the message was received, until the included \texttt{max-age} expiration. This is similar in spirit to the response header for the HTTP Strict Transport Security (HSTS) mechanism.

In the opportunistic mode, the user's path policy is interpreted as a preference. If a website is available via SCION but no policy-compliant path is available, e.g., if the user navigates to a website that is hosted in an AS that the user indicated to avoid, the website will still load. The user is informed of the non-compliance with the same indicator as when loading the website via legacy IP.
Strict mode only allows policy-compliant paths and the browser will display
a connection error if no such path is found.

\subsection{SCION Detection for Domains}\label{subsec:detection}

Since SCION uses a different address scheme (a combination of SCION ISD, AS and local IPv4/6 address) than IP, adapting address resolution is required. While keeping a curated list of SCION-available domains in the browser provides a reasonable starting point, this approach does not scale. Consequently, we provide two dynamic approaches for detecting SCION-accessible domains.

First, the presence of the \texttt{Strict-SCION} header can also be used as an indicator to advertise the availability of SCION to users who have not yet enabled SCION support in their browser -- similarly to the \texttt{Onion-Location} header~\cite{onion-location-header} that advertises the availability of Tor in the Brave browser \cite{bravetorheader}.

Secondly, additional TXT records indicating a SCION address can be configured in existing DNS records. The HTTP proxy can determine to use SCION, or to fall back to IP if no SCION address is available.

\section{Implementation and Evaluation}
\label{sec:impl-and-eval}

We implement a SCION-based PAN architecture prototype as a browser extension and deploy it in the Brave browser. The implementation of the SCION transport is included in a separate \emph{HTTP proxy} local process.

\subsection{Prototype Implementation}
Our prototype implements the concept presented in Section \ref{sec:pan-browsing}. As discussed in Section \ref{sec:network-innovations-in-browser}, the limited functionality of the WebExtensions API requires our prototype to proxy all of the browser's requests. First, the browser extension uses specific API calls to the HTTP proxy to apply path policies chosen by users. Second, the extension configures the default proxy for all network requests to the HTTP proxy component, which then decides on using either SCION or IPv4/6. Note that integrating parts of the extension into the browser code could overcome this limitation.

The browser extension itself, i.e., the JavaScript logic, has two roles. First, it presents the options and settings in the browser's user interface and configures the proxy component according to the user's preferences. Furthermore, it takes care of implementing the \emph{strict mode}; as the proxy is a regular HTTP proxy it does not have the necessary context to decide whether strict mode should be enabled for a particular request or not. When the extension determines that a request is to be performed in strict mode, it first checks whether the resource is available via a policy-compliant SCION path. If there is such a path, the request is forwarded via the proxy, otherwise the request is blocked.

Since SCION local AS communication is based on UDP, SCION-aware applications can operate without OS support, simplifying the deployment of our HTTP proxy. We exclusively use QUIC as the transport layer for all web traffic over SCION. For HTTP/1 and HTTP/2, we map the TCP data stream into a single bidirectional QUIC stream. 
The primary reason for this deviation from the norm is the stated goal to implement all of this inside the application -- high-quality library implementations of QUIC which can be made to work on top of SCION are readily available.
Our implementation is based on the \emph{quic-go}~library~\cite{quic-go} as well as Go's built-in HTTP implementation.

To complement our client-side implementation, we have implemented a simple reverse proxy to add SCION support to web servers. Alternatively, Go-based web servers can be compiled with our PAN library to include SCION support directly. This could also be used to build a SCION module for popular Go-based web servers like Caddy or Traefik. 

\subsection{Experimental Results}

\begin{figure}[t]
\begin{center}
\includegraphics[width=6.7cm]{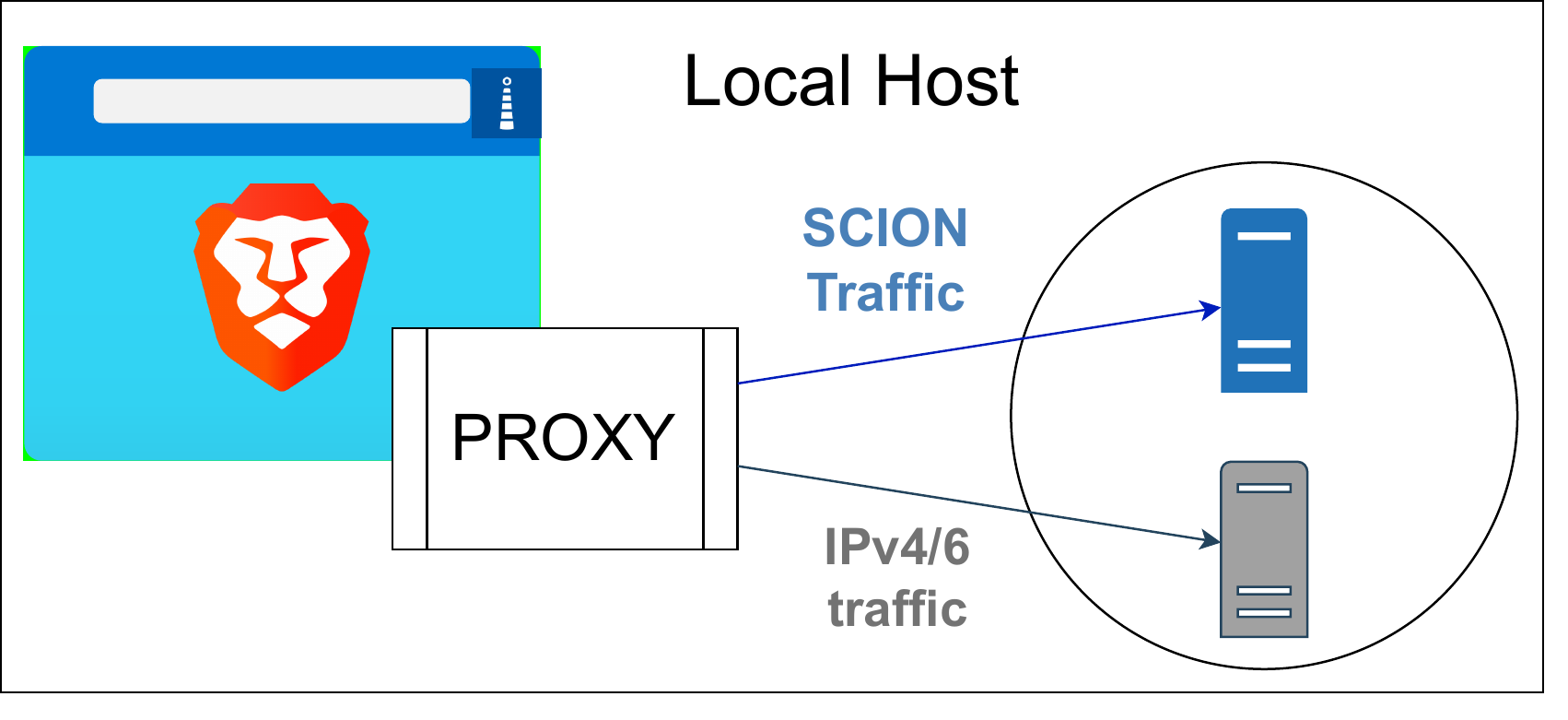}
\caption{Local setup. The blue host is a SCION-enabled file server and the grey one is a TCP/IP file server without SCION connectivity.}
\label{fig:local-setup}
\end{center}
\end{figure}

\begin{figure}[t]
\includegraphics[clip=true,width=7.5cm]{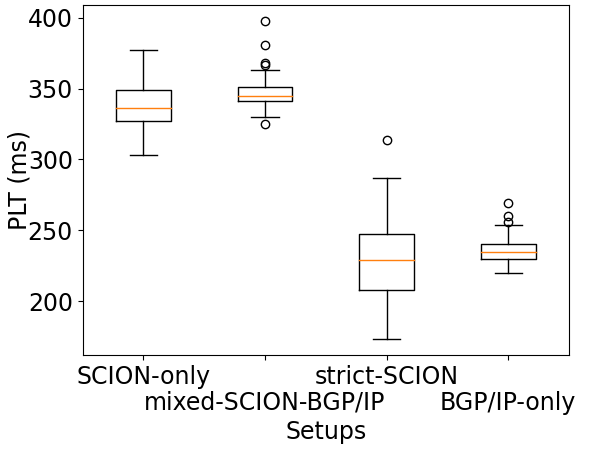}
\caption{Page Load Time (ms) for each experiment type.}
\label{fig:boxplot-local}
\end{figure}

\begin{figure}[t]
\includegraphics[width=\linewidth]{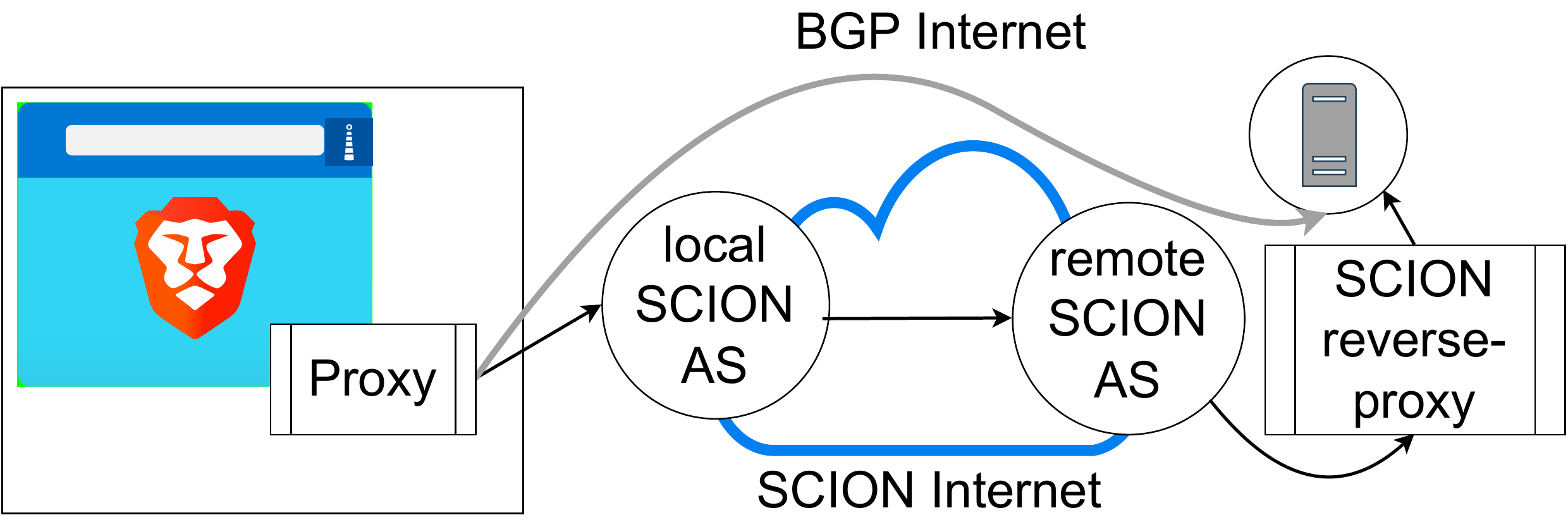}
\caption{Remote setup. The browser loads the SCION extension; the grey host represents a TCP/IP server that is also reachable over a nearby SCION reverse proxy.}
\label{fig:remote}
\end{figure}

\begin{figure}[t]
\includegraphics[width=.95\linewidth]{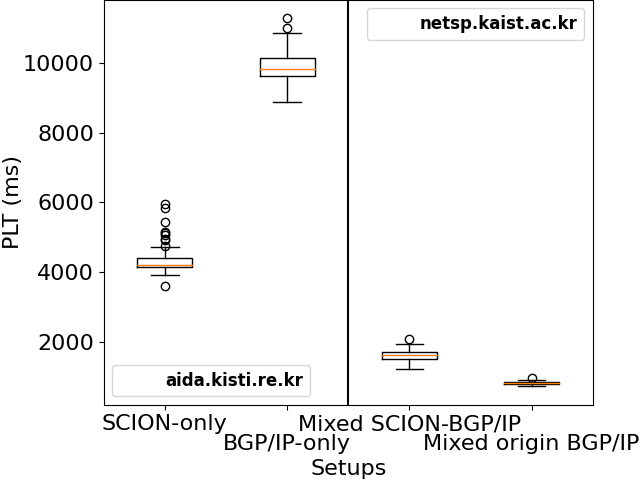}
\caption{Comparison of Page Load Time (ms) for remote pages over SCION and IPv4/6, containing either resources from one or multiple origins.}

\label{fig:boxplot-gateway}
\end{figure}

\begin{figure}[t]
\includegraphics[width=.95\linewidth]{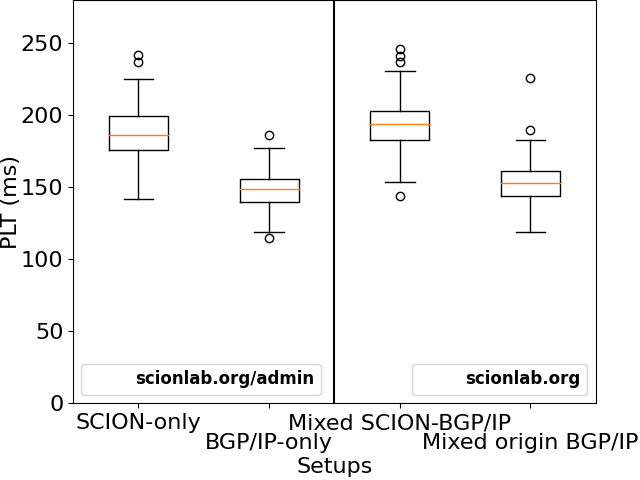}
\caption{Comparison of Page Load Time (ms) for an AS local page over SCION and IPv4/6, containing either resources from one or multiple origins.}

\label{fig:boxplot-gateway-AS-local}
\end{figure}

We present our experimental results obtained from two different setups: a local environment, where the file servers providing static web content run on the same host as the Brave browser and HTTP proxy, and a distributed setup, where we access remote web servers with a SCION reverse proxy.

The local setup shown in Figure~\ref{fig:local-setup} is hosted on a laptop, where no requests exit that system. We use two file servers providing static content hosted on a VM: the SCION File Server (blue host) provides content over SCION, while the TCP/IP File Server (grey host) provides resources over TCP/IP. The HTTP proxy can establish connections both to SCION and TCP/IP servers. These experiments compare the Page Load Time (PLT) running the extension compared to the PLT for the standard browsing experience.

The box plots in Figure~\ref{fig:boxplot-local} depict four experiments. 
The \textit{SCION-only} experiment shows the load time for a static website in which all resources are located on the SCION FS.
In the \textit{mixed SCION-IP} experiment, the HTTP proxy fetches resources from both servers. 
In the \textit{strict-SCION} experiment, the browser extension runs in \texttt{Strict-SCION} mode,

thus only requesting SCION resources and blocking all others. In this experiment, only one resource is located on the SCION FS, while the others are located on the TCP/IP FS and are thus blocked. 
Finally, the \textit{BGP/IP-Only} experiment shows the PLT for the browsing experience with the extension disabled, i.e., requests are not intercepted by the extension and do not traverse the HTTP proxy. 
The results show a longer PLT for the \textit{SCION-only} and the \textit{mixed SCION-IP} (approximately 100 ms) with respect to the PLT when the extension is disabled (\textit{BGP/IP-Only}) and to the \textit{strict-SCION} experiment. In the first two experiments, the requests and responses go through the extension and the HTTP proxy for all the resources creating overhead to the communication. In \textit{strict-SCION}, only the request to one resource is forwarded, while the rest are blocked, thus shortening the PLT. In the \textit{BGP/IP-Only} experiment, the extension is fully disabled, thus, the overhead is removed. With tighter SCION integration in the browser and web server, we expect the overhead to disappear, and even to achieve better performance for large objects with performance-based path selection.

The distributed setup is depicted in Figure~\ref{fig:remote}. For this setup, we carry out two different experiments. The first experiment, in Figure~\ref{fig:boxplot-gateway}, shows the PLT for resources hosted in distant locations. For the single origin page, we observe that the PLT improves significantly when the resource is loaded via SCION. The reason is that, in this particular case, SCION benefits from path awareness to pick a lower-latency path than the one traversed by the legacy IPv4/6 traffic. The second experiment, in Figure~\ref{fig:boxplot-gateway-AS-local}, shows the PLT for websites located closer to the browser host. Similarly to the local setup, we observe that when paths are similar, the extension adds a small overhead compared to the baseline.

\section{Related Work}

We discuss related work in the general areas of next-gene\-ra\-tion Internet architectures with a focus on innovation, app\-li\-cation-level network APIs supporting advanced networking functionality, and browser-based network innovations.

Several next-generation Internet projects had the aim to facilitate network innovation~\cite{McGeer2016,bavier2006vini,Crowcroft2003,Han12xia,Jacobson09ccn,RaNaVe2012,KSBFGGGMPRRAK2011,GoGaShSt2009,Rouskas2013,route_bazaar_2015}.

We argue that browser-based implementation is an effective means for rapid transition of these innovations, without the need for OS support. A contrary view for the application-level API is by the RINA architecture, which argues that the network-level API should be similar in simplicity to local inter-process communication~\cite{Wang_rina2014}. As we argue in this paper, some network properties require user input, and thus need to be exposed to the application. Network testbeds built to support research and innovation provide advanced functionalities~\cite{McGeer2016,bavier2006vini,PlanetLab,kwon2020scionlab}, which can be used to study the API to the OS and applications.

The Transport Services Application Programming Interface (TAPS) is a new network API, built to facilitate application's use of new functionality~\cite{RFC8303,RFC8095,draft-ietf-taps-arch-12}. TAPS provides a generic interface also for PAN functionality~\cite{PANAPI}, which applications can make use of for achieving the properties we discuss in this paper.

Several low-level APIs exist for high-speed packet processing, which bypass the OS, for instance DPDK~\cite{linux-dpdk}, AF-XDP~\cite{hoilandjorgensen2018express}, VPP IO~\cite{vpp-io}, and PF\_RING~\cite{pf_ring}.

Such frameworks can be used to achieve high-speed PAN communication without dedicated kernel support.

Browser-based network innovation has already occurred several times: QUIC implementation in browsers~\cite{quic,quic-firefox,quic-opera} and servers~\cite{quic-caddy, quic-nginx}, 
WebRTC enabled cross-browser video-conferencing using standard APIs without OS support or additional plugins~\cite{webrtc-interop}, and VPNs in browsers~\cite{firefox-vpn,brave-vpn,opera-vpn,apple-cloud-relay-vpn,edge-vpn}. These highly successful deployments demonstrate the power of browser-led innovation for the deployment of new network functionality.

\section{Conclusion}

We propose a design, an open-source prototype implementation \cite{scion-apps-skip, scion-browser-extensions}, and initial experiments of integrating path-aware networking into the Brave browser. To fully unlock the benefits of PAN architectures, both web servers and browsers should be equipped with the ability to perform context-dependent path selection.
On the side of web servers, much engineering work remains to enable PAN properties directly on the server, without the need for a reverse proxy. While our prototype implementation is restricted by the limitations of browser extensions, we plan to move features into browser code to fully capture the potential of PAN in browsers, and bundle the SCION setup in a single executable for simplified deployment. Another direction is to implement further path policies, i.e., optimizing network paths for energy, or CO$_2$ footprint.

Another interesting future direction is to enable path negotiation between the server and the browser, enabling another dimension of achievable properties and innovation.

Network innovation has traditionally been driven from the OS. Thanks to recent developments, we observe that novel, feature-rich browsers provide a potent approach to drive innovation for advanced networking mechanisms -- both out of necessity (user involvement) and simplicity (ease of deployment and use for the average user). Some properties cannot be meaningfully decided in the network or network layer, but instead depend on user preferences. Consequently, the browser provides a perfect environment for such properties, and an exciting platform for network innovation.

\bibliographystyle{abbrv} 
\bibliography{hotnets22}

\begin{thebibliography}{10}

\bibitem{pathlanguage}
{Anapaya Systems}.
\newblock {Path Policy Language Design}.
\newblock \url{https://scion.docs.anapaya.net/en/latest/PathPolicy.html}, 2022.

\bibitem{apple-cloud-relay-vpn}
Apple.
\newblock {About iCloud Private Relay}.
\newblock \url{https://support.apple.com/en-us/HT212614}, 2021.

\bibitem{bavier2006vini}
A.~Bavier, N.~Feamster, M.~Huang, L.~Peterson, and J.~Rexford.
\newblock In {VINI} veritas: realistic and controlled network experimentation.
\newblock In {\em Proceedings of the 2006 conference on applications,
  technologies, architectures, and protocols for computer communications
  (SIGCOMM)}, 2006.

\bibitem{brave-internal-stats}
Brave.
\newblock Internal usage stats.
\newblock Obtained via direct communication, 2022.

\bibitem{brave-mau}
Brave.
\newblock Platform stats.
\newblock \url{https://brave.com/transparency/}, 2022.

\bibitem{brave-tor}
{Brave Software}.
\newblock Brave introduces beta of private tabs with tor for enhanced privacy
  while browsing.
\newblock \url{https://brave.com/tor-tabs-beta/}, 2018.

\bibitem{brave-vpn}
{Brave Software}.
\newblock {Brave Firewall + VPN}.
\newblock \url{https://brave.com/firewall-vpn/}, 2020.

\bibitem{bravetorheader}
{Brave Team}.
\newblock {Handle onion-location HTTP header \& .onion domain}.
\newblock \url{https://github.com/brave/brave- core/pull/6762}, 2022.

\bibitem{route_bazaar_2015}
I.~Castro, A.~Panda, B.~Raghavan, S.~Shenker, and S.~Gorinsky.
\newblock {Route Bazaar}: Automatic interdomain contract negotiation.
\newblock In {\em Proceedings of USENIX Conference on Hot Topics in Operating
  Systems (HotOS)}, 2015.

\bibitem{Chuat2022}
L.~Chuat, M.~Legner, D.~Basin, D.~Hausheer, S.~Hitz, P.~Müller, and A.~Perrig.
\newblock {\em The Complete Guide to {SCION}}.
\newblock Springer International Publishing, 2022.

\bibitem{quic-go}
L.~Clemente.
\newblock quic-go.
\newblock \url{https://github.com/lucas-clemente/} \url{quic-go}, 2022.

\bibitem{quic-nginx}
L.~Crilly.
\newblock {Introducing a Technology Preview of NGINX Support for QUIC and
  HTTP/3}.
\newblock
  \url{https://www.nginx.com/blog/introducing-technology-preview-nginx-}
  \url{support-for-quic-http-3/}, 2020.

\bibitem{Crowcroft2003}
J.~Crowcroft, S.~Hand, R.~Mortier, T.~Roscoe, and A.~Warfield.
\newblock {Plutarch: An Argument for Network Pluralism}.
\newblock {\em ACM SIGCOMM Computer Communication Review}, 33(4):258–266,
  Aug. 2003.

\bibitem{quic-firefox}
D.~Damjanovic.
\newblock {QUIC and HTTP/3 Support now in Firefox Nightly and Beta}.
\newblock
  \url{https://hacks.mozilla.org/2021/04/quic-and-http-3-support-now-in-firefox-}
  \url{nightly-and-beta/}, 2021.

\bibitem{USENIX:DinMatSyv04}
R.~Dingledine, N.~Mathewson, and P.~Syverson.
\newblock Tor: The {Second-Generation} onion router.
\newblock In {\em USENIX Security Symposium}, Aug. 2004.

\bibitem{RFC8095}
G.~Fairhurst, B.~Trammell, and M.~K{\"u}hlewind.
\newblock {Services Provided by IETF Transport Protocols and Congestion Control
  Mechanisms}.
\newblock RFC 8095, IETF, Jan. 2020.

\bibitem{RFC8402}
C.~Filsfils, S.~Previdi, L.~Ginsberg, B.~Decraene, S.~Litkowski, and R.~Shakir.
\newblock {Segment Routing Architecture}.
\newblock RFC 8402, IETF, July 2018.

\bibitem{conext2021colibri}
G.~Giuliari, D.~Roos, M.~Wyss, J.~A. García-Pardo, M.~Legner, and A.~Perrig.
\newblock Colibri: A cooperative lightweight inter-domain bandwidth-reservation
  infrastructure.
\newblock In {\em Proceedings of ACM International Conference on emerging
  Networking EXperiments and Technologies (CoNEXT)}, Dec. 2021.

\bibitem{GoGaShSt2009}
P.~B. Godfrey, I.~Ganichev, S.~Shenker, and I.~Stoica.
\newblock Pathlet routing.
\newblock In {\em Proceedings of ACM SIGCOMM Conference on Data Communication},
  2009.

\bibitem{Han12xia}
D.~Han, A.~Anand, F.~Dogar, B.~Li, H.~Lim, M.~Machado, A.~Mukundan, W.~Wu,
  A.~Akella, D.~G. Andersen, J.~W. Byers, S.~Seshan, and P.~Steenkiste.
\newblock {XIA}: Efficient support for evolvable internetworking.
\newblock In {\em Proceedings of USENIX Symposium on Networked Systems Design
  and Implementation (NSDI)}, 2012.

\bibitem{hoilandjorgensen2018express}
T.~H\o{}iland-J\o{}rgensen, J.~D. Brouer, D.~Borkmann, J.~Fastabend,
  T.~Herbert, D.~Ahern, and D.~Miller.
\newblock {The EXpress Data Path}: Fast programmable packet processing in the
  operating system kernel.
\newblock In {\em Proceedings of ACM International Conference on Emerging
  Networking EXperiments and Technologies}, CoNEXT, 2018.

\bibitem{quic-caddy}
M.~Holt.
\newblock caddyserver/caddy release note 0.9.
\newblock \url{https://github.com/caddyserver/caddy/releases/tag/v0.9.0}, 2016.

\bibitem{webrtc-interop}
P.~Höglund.
\newblock {Chrome - Firefox WebRTC Interop Test}.
\newblock \url{https://testing.googleblog.com/}
  \url{2014/08/chrome-firefox-webrtc-interop-} \url{test-pt-1.html}, 2014.

\bibitem{Jacobson09ccn}
V.~Jacobson, D.~K. Smetters, J.~D. Thornton, M.~F. Plass, N.~H. Briggs, and
  R.~L. Braynard.
\newblock Networking named content.
\newblock In {\em Proceedings of International Conference on Emerging
  Networking Experiments and Technologies (CoNEXT)}, 2009.

\bibitem{onion-location-header}
G.~Kadianakis.
\newblock Onion-location.
\newblock
  \url{https://community.torproject.org/onion-services/advanced/onion-location/},
  2018.

\bibitem{KSBFGGGMPRRAK2011}
T.~Koponen, S.~Shenker, H.~Balakrishnan, N.~Feamster, I.~Ganichev, A.~Ghodsi,
  P.~B. Godfrey, N.~McKeown, G.~Parulkar, B.~Raghavan, J.~Rexford, S.~Arianfar,
  and D.~Kuptsov.
\newblock Architecting for innovation.
\newblock {\em ACM SIGCOMM Computer Communication Review}, July 2011.

\bibitem{conext2021deployment}
C.~Kr\"{a}henb\"{u}hl, S.~Tabaeiaghdaei, C.~Gloor, J.~Kwon, A.~Perrig,
  D.~Hausheer, and D.~Roos.
\newblock Deployment and scalability of an inter-domain multi-path routing
  infrastructure.
\newblock In {\em Proceedings of ACM International Conference on emerging
  Networking EXperiments and Technologies (CoNEXT)}, Dec. 2021.

\bibitem{PANAPI}
T.~Kr\"{u}ger and D.~Hausheer.
\newblock {Towards an API for the Path-Aware Internet}.
\newblock In {\em Proceedings of ACM SIGCOMM Workshop on Network-Application
  Integration (NAI)}, 2021.

\bibitem{kwon2020scionlab}
J.~Kwon, J.~A. Garc{\'i}a-Pardo, M.~Legner, F.~Wirz, M.~Frei, D.~Hausheer, and
  A.~Perrig.
\newblock {SCIONLab}: A next-generation {Internet} testbed.
\newblock In {\em Proceedings of IEEE Conference on Network Protocols (ICNP)},
  2020.

\bibitem{alibi2015}
D.~Levin, Y.~Lee, L.~Valenta, Z.~Li, V.~Lai, C.~Lumezanu, N.~Spring, and
  B.~Bhattacharjee.
\newblock Alibi routing.
\newblock In {\em Proceedings of ACM Conference on Special Interest Group on
  Data Communication (SIGCOMM)}, Aug. 2015.

\bibitem{linux-dpdk}
{Linux Foundation}.
\newblock {Data Plane Development Kit}.
\newblock \url{https://www.dpdk.org/}, 2018.

\bibitem{McGeer2016}
R.~McGeer, M.~Berman, C.~Elliott, and R.~Ricci, editors.
\newblock {\em The {GENI} Book}.
\newblock Springer International Publishing, 2016.

\bibitem{edge-vpn}
Microsoft.
\newblock {Introducing Microsoft Edge Secure Network}.
\newblock
  \url{https://techcommunity.microsoft.com/t5/articles/introducing-microsoft-edge-secure-network/m-p/3367243},
  2022.

\bibitem{chromium-release-cycle}
A.~Mineer.
\newblock Speeding up chrome's release cycle.
\newblock
  \url{https://blog.chromium.org/2021/03/speeding-up-release-cycle.html}, 2021.

\bibitem{firefox-vpn}
Mozilla.
\newblock {Mozilla VPN: Protect Your Entire Device}.
\newblock \url{https://www.mozilla.org/en-US/products/vpn/}, 2020.

\bibitem{scion-apps-skip}
{Netsec ETHZ}.
\newblock {Skip Proxy}.
\newblock \url{https://github.com/netsec-ethz/scion-apps/tree/master/skip},
  2022.

\bibitem{scion-browser-extensions}
{Netsec ETHZ, OVGU Magdeburg}.
\newblock {SCION Browser Extensions}.
\newblock \url{https://github.com/netsys-lab/scion-browser-extensions}, 2022.

\bibitem{pf_ring}
{NTOP}.
\newblock {PF\_RING}.
\newblock \url{https://github.com/ntop/PF\_RING}, 2022.

\bibitem{opera-vpn}
Opera.
\newblock {Opera Browser with free VPN}.
\newblock \url{https://www.opera.com/features/free-vpn}, 2016.

\bibitem{quic-opera}
{Opera Team}.
\newblock {Changelog 41.0.2340 - Opera Desktop}.
\newblock \url{https://blogs.opera.com/desktop/changelog-opera-41/}, 2016.

\bibitem{draft-ietf-taps-arch-12}
T.~Pauly, B.~Trammell, A.~Brunstrom, G.~Fairhurst, C.~Perkins, P.~S. Tiesel,
  and C.~A. Wood.
\newblock {An Architecture for Transport Services}.
\newblock draft draft-ietf-taps-arch-12, IETF, Jan. 2022.

\bibitem{RaNaVe2012}
D.~Raychaudhuri, K.~Nagaraja, and A.~Venkataramani.
\newblock {MobilityFirst}: A robust and trustworthy mobility-centric
  architecture for the future {Internet}.
\newblock {\em ACM SIGMOBILE Mobile Computing and Communications Review}, July
  2012.

\bibitem{Rouskas2013}
G.~Rouskas, I.~Baldine, K.~Calvert, R.~Dutta, J.~Griffioen, A.~Nagurney, and
  T.~Wolf.
\newblock {ChoiceNet}: Network innovation through choice.
\newblock In {\em Proceedings of International Conference on Optical Networking
  Design and Modeling (ONDM)}, 2013.

\bibitem{web-browser-market-share}
Statista.
\newblock Market share of leading internet browsers.
\newblock \url{https://www.statista.com/statistics/}
  \url{276738/worldwide-and-us-market-}
  \url{share-of-leading-internet-browsers/}, Feb 21 2022.

\bibitem{web-browser-stats}
Statista.
\newblock Web browsers - statistics \& facts.
\newblock \url{https://www.statista.com/topics/5684/web-browsers/}, Feb 7 2022.

\bibitem{quic}
{The Chromium Project}.
\newblock {QUIC, a multiplexed transport over UDP}.
\newblock \url{https://www.chromium.org/quic/#History}, 2022.

\bibitem{vpp-io}
{The Fast Data Project (FD.io)}.
\newblock {VPP/What is VPP?}
\newblock \url{https://wiki.fd.io/view/VPP/} \url{What\_is\_VPP\%3F}, 2017.

\bibitem{PlanetLab}
{The PlanetLab Consortium}.
\newblock {PlanetLab}, an open platform for developing, deploying, and
  accessing planetary-scale services.
\newblock \url{https://www.planet-lab.org/}, 2016.

\bibitem{tor-usage}
{Tor Project}.
\newblock User metrics.
\newblock
  \url{https://metrics.torproject.org/userstats-relay-country.html?start=2011-01-01},
  21 June 2022.

\bibitem{pan2018}
B.~Trammell, J.-P. Smith, and A.~Perrig.
\newblock Adding path awareness to the internet architecture.
\newblock {\em IEEE Internet Computing}, 22(2):96--102, Mar. 2018.

\bibitem{Wang_rina2014}
Y.~Wang, I.~Matta, F.~Esposito, and J.~Day.
\newblock {Introducing ProtoRINA: A Prototype for Programming
  Recursive-Networking Policies}.
\newblock In {\em Proceedings of ACM SIGCOMM Conference}, volume~44, page
  129–131, New York, NY, USA, jul 2014. Association for Computing Machinery.

\bibitem{RFC8303}
M.~Welzl, M.~T{\"u}xen, and N.~Khademi.
\newblock {On the Usage of Transport Features Provided by IETF Transport
  Protocols}.
\newblock RFC 8303, IETF, Dec. 2020.

\end{thebibliography}

\end{document}